\documentclass[manuscript]{acmart}
\AtBeginDocument{%
  \providecommand\BibTeX{{%
    \normalfont B\kern-0.5em{\scshape i\kern-0.25em b}\kern-0.8em\TeX}}}

\setcopyright{none}
\copyrightyear{2024}
\acmYear{2024}

\acmConference[ASSETS '24 Workshop: The Future of Urban Accessibility]{}{October 2024}{Virtual}

\usepackage{xpatch}

\makeatletter
\xpatchcmd{\ps@firstpagestyle}{Manuscript submitted to ACM}{}{\typeout{First patch succeeded}}{\typeout{first patch failed}}
\xpatchcmd{\ps@standardpagestyle}{Manuscript submitted to ACM}{}{\typeout{Second patch succeeded}}{\typeout{Second patch failed}}    \@ACM@manuscriptfalse
\makeatother

\acmBooktitle{Proceedings of The Future of Urban Accessibility: The Role of AI Workshop at The 26th International ACM SIGACCESS
Conference on Computers and Accessibility (ASSETS ’24)} 
\acmISBN{}

\usepackage{subfigure}

\usepackage{appendix}

\usepackage{subfigure}

\usepackage{graphicx}
\usepackage{multirow}
\usepackage{pifont}
\usepackage{lscape}
\usepackage{array}

\usepackage{mathtools}

\begin{document}

\title[Making Urban Art Accessible: Current Art Access Techniques, \\Design Considerations, and the Role of AI]{Making Urban Art Accessible: Current Art Access Techniques, Design Considerations, and the Role of AI}

\author{Lucy Jiang}
\affiliation{%
  \institution{Human Centered Design and Engineering, University of Washington}
  \city{Seattle, WA}
  \country{USA}}
\email{lucjia@uw.edu}

\author{Jon E. Froehlich}
\affiliation{%
  \institution{Paul G. Allen School of Computer Science \& Engineering, University of Washington}
  \city{Seattle, WA}
  \country{USA}}
\email{jonf@cs.uw.edu}

\author{Leah Findlater}
\affiliation{%
  \institution{Human Centered Design and Engineering, University of Washington}
  \city{Seattle, WA}
  \country{USA}}
\email{leahkf@uw.edu}


\begin{abstract}
Public artwork, from vibrant wall murals to captivating sculptures, can enhance the aesthetic of urban spaces, foster a sense of community and cultural identity, and help attract visitors. Despite its benefits, most public art is visual, making it often inaccessible to blind and low vision (BLV) people. In this workshop paper, we first draw on art literature to help define the space of public art, identify key differences with curated art shown in museums or galleries, and discuss implications for accessibility. We then enumerate how existing art accessibility techniques may (or may not) transfer to urban art spaces. We close by presenting future research directions and reflecting on the growing role of AI in making art accessible. 

\end{abstract}



\begin{teaserfigure}
  \includegraphics[width=\textwidth]{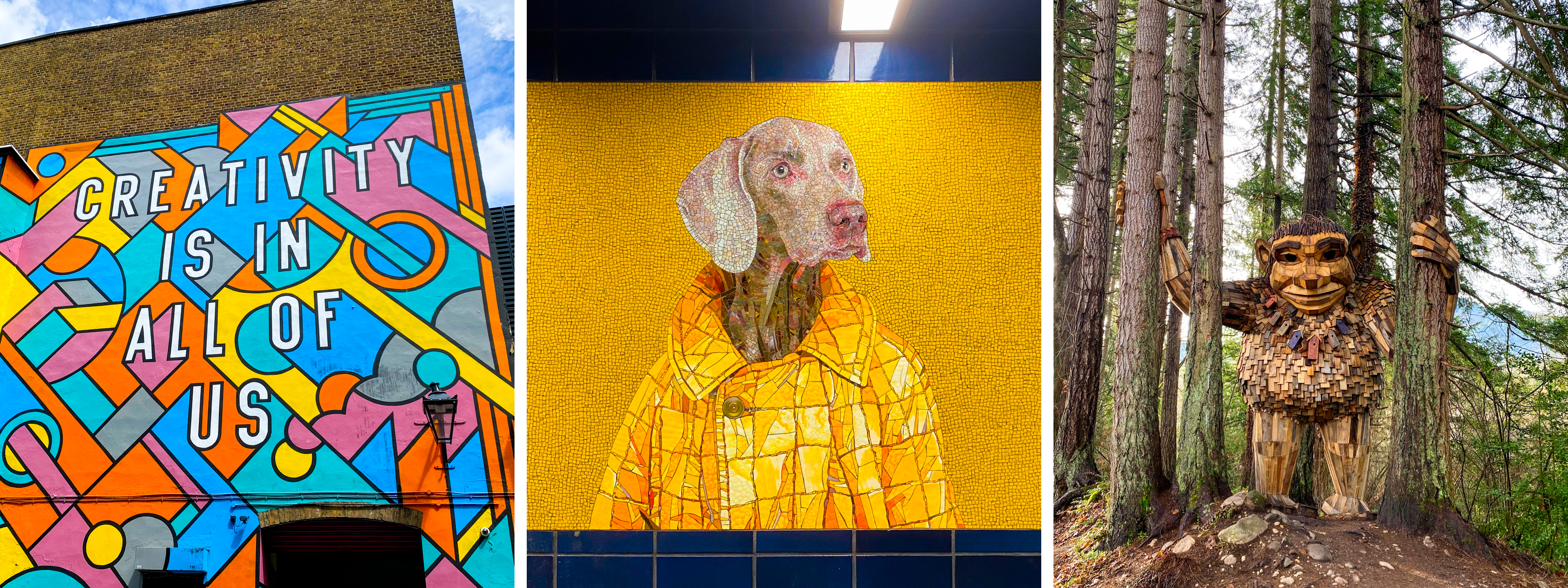}
  \caption{Three different examples of public art. From left to right: a large mural titled \textit{Creativity Is in All of Us} by Rob Lowe (also known as Supermundane) in London’s Covent Garden, a large mosaic from the \textit{Stationary Figures} art series by William Wegman on the wall of a New York City subway station, and the Jakob Two Trees troll sculpture by Thomas Dambo in Issaquah, Washington.}
  \Description{Three images from left to right, showing a mural, mosaic, and sculpture. The mural features brightly colored overlapping polygons in the background and reads “CREATIVITY IS IN ALL OF US” in white capital letters. The mosaic shows a brown-gray Weimaraner dog looking off to the right and wearing a yellow quilted jacket. The sculpture is a 14-foot tall wooden troll, wearing a necklace made of birdhouses and posing with its two arms wrapped around tall trees next to it.}
  \label{fig:teaser}
\end{teaserfigure}

\maketitle

\section{Introduction}
Public artworks, such as murals, installations, sculptures, graffiti tags, and more, are ubiquitous in today’s urban environments. In the city of Seattle, Washington, there are more than 400 permanent public art pieces and over 3,000 portable or temporary works, intended to \textit{“simultaneously enrich citizens’ daily lives and give voice to artists”} \cite{seattlepublicart}. Distinct from art produced for museums or galleries, public art is interwoven into the culture of an urban community, playing a crucial role in fostering cultural identity, enhancing aesthetic appeal, and even improving economic outcomes. 

Despite its benefits, most art is highly visual and often inaccessible to blind and low vision (BLV) people. There is an emerging body of research investigating the accessibility of art in museum and gallery settings (e.g., \cite{li2023understanding, rector2017eyes, cavazos2018interactive, cavazos2021accessible, cavazos2021multi, holloway2019making, kwon2022supporting}) and in one’s everyday personal life (e.g., \cite{chheda2024engaging}). Prior work includes a variety of approaches to improve nonvisual access to artwork, such as audio descriptions \cite{kwon2022supporting}, tactile graphics \cite{cavazos2018interactive}, and proxemic audio interfaces \cite{rector2017eyes}. However, those approaches are typically designed for art exploration in controlled museum environments, where art is carefully curated, the surroundings are generally quiet, there are clear pathways for moving from one artwork to another, and artists often provide additional blurbs or information about their work. Public art, on the other hand, can be situated in a wide range of settings, including chaotic and noisy intersections or tranquil public parks. Furthermore, it may be either obvious (e.g., a full wall mural) or inconspicuous (e.g., a graffiti tag on a parking meter).

There is limited understanding on how to make public art more accessible to BLV people. Thus, our research questions are exploratory: 
\begin{itemize}	
	\item What are the potential scope and characteristics of public art accessibility?
	\item How and to what extent may art accessibility techniques developed for more controlled settings (e.g., museums) translate to public art consumption?
	\item How can technology aid in making public art more accessible?
\end{itemize}

To address these questions, we first define the space of urban art and accessibility, extending discourse on urban accessibility for blind and low vision people beyond orientation and navigation. Then, we review the broader scope of art and image accessibility work, synthesize existing techniques, and discuss how art accessibility techniques may be transferable between private and public settings. Lastly, we share design considerations and propose future research directions in this space.

\section{Defining the Space of Urban Art and Accessibility}
What is public art? Common conceptions of public art include murals artistically presenting a city’s name or sculptures recognizing famous historical figures. However, public art is more than just museum artwork in an urban setting. In \textit{Mapping the Terrain: New Genre Public Art}, public art is defined as not only sculptures, installations, or visual arts in public spaces \textemdash{} instead, they are medium-agnostic artworks designed to \textit{“communicate and interact with a broad and diversified audience about issues directly relevant to their lives”} \cite{lacy1995mapping}. Public art can be any type of artistic expression that engages with people and places to foster conversation and deepen meaning. For this workshop paper, we scope the space of urban art and accessibility to three popular \textit{families} of public art: murals, mosaics, and sculptures.

\begin{figure*}[b!]
	\centering
	\includegraphics[width=0.6\textwidth]{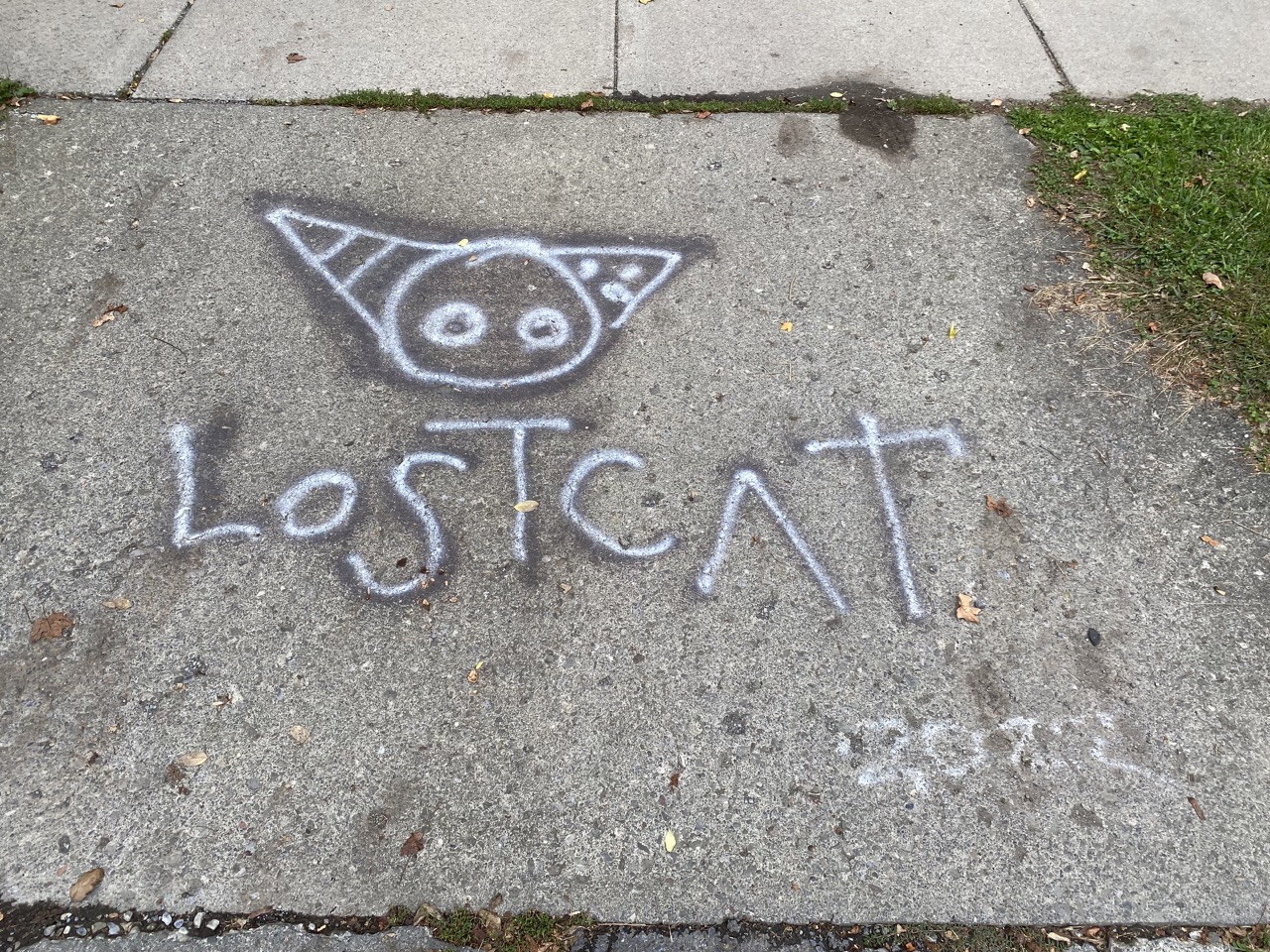}
	\caption{One of many \textit{Lost Cat} graffiti tags in Ithaca, New York. The artist remained anonymous until his passing.}
	\Description{A graffiti tag made with white spray paint on pavement. On top, there is a cartoonish outline of a cat’s head with two triangles for ears and a slightly flattened oval for the head. The left ear has two lines inside of it, the right ear has three dots inside of it, and the cat’s eyes are two small circles. Underneath reads “LoSTcAt 2022” in a mixture of upper- and lowercase handwriting.}
	\label{fig:mural}
\end{figure*}

\textbf{Murals} are typically painted on flat surfaces, such as walls or streets, and generally do not include any tactile or three-dimensional elements. Some examples of well-known murals include Banksy’s \textit{Love Is In The Air} which criticizes militarism and war \cite{sothebys} or Keith Haring’s many subway paintings. Drawing on work by Hughes \cite{hughes2009street}, we consider graffiti art (including tags) to be a subset of mural-type street art, and recognize its artistic, identity-centric, and community building merits. One such example is \textit{Lost Cat}, a prominent and well-loved graffiti tag in the Ithaca, New York area (Figure \ref{fig:mural}). Given that these artworks are not inherently tactile, however, they are often inaccessible without visual descriptions.

\begin{figure*}[h!]
	\centering
	\includegraphics[width=0.6\textwidth]{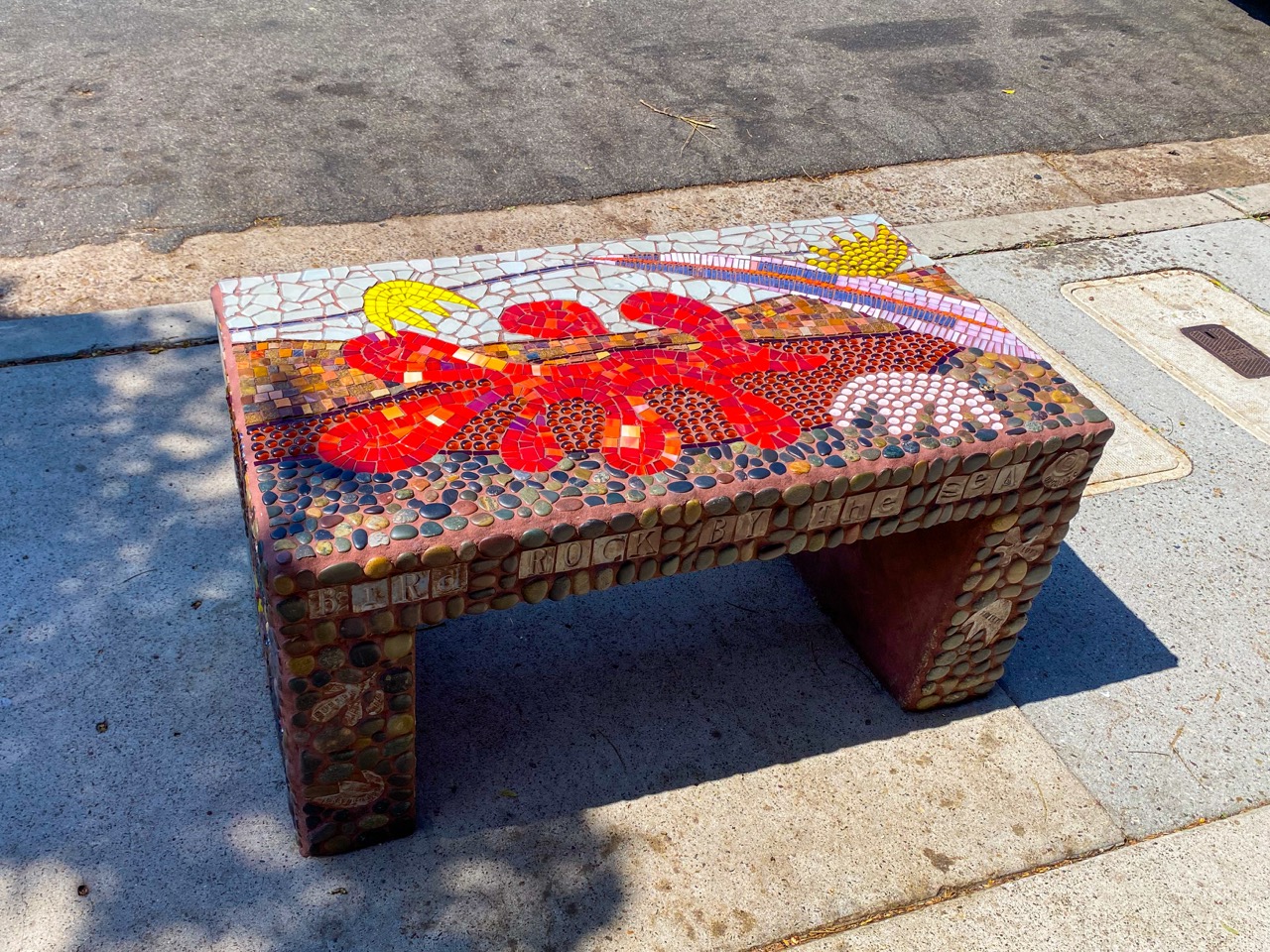}
	\caption{A bench in the \textit{Benches of Bird Rock} series, which is covered in a variety of colorful ceramic tiles, stones, and marbles in San Diego, California. The mosaics were created by Jane Wheeler.}
	\Description{A small, rectangular bench with a mosaic on the top. Similar to the style of artist Henri Matisse’s collages, the mosaic features a large red leaf-like shape in the center, created with small and flat square tiles. Above the red portion, there is a white backdrop created with irregular opaque glass pieces. Between the white and red sections, there are two small shapes created with yellow tiles, resembling half of a cracked eggshell. Beneath the red shape, there is a small section created with amber glass marbles, and the bottom of the mosaic is filled in with small stones. The side of the bench reads “BiRd ROCK BY The seA” in mixed upper- and lowercase.}
	\label{fig:mosaic}
\end{figure*}

\textbf{Mosaics} are made of stone, tile, glass, or clay. They are often tactile and three-dimensional, though this may vary depending on the type, thickness, or texture of the materials. Many New York City subway stations feature elaborate mosaics for riders to appreciate during transit. Another example of tactile public art is benches (Figure \ref{fig:mosaic}), which blend form and function by utilizing various materials to create a textured surface on commonplace infrastructure. While mosaics have tactile elements, they may still be inaccessible if the scale of the art is too large or too small to comprehend via touch, if the tactile experience is too uniform, or if they are located in places that are hard to physically access.

\begin{figure*}[]
	\centering
	\includegraphics[width=0.6\textwidth]{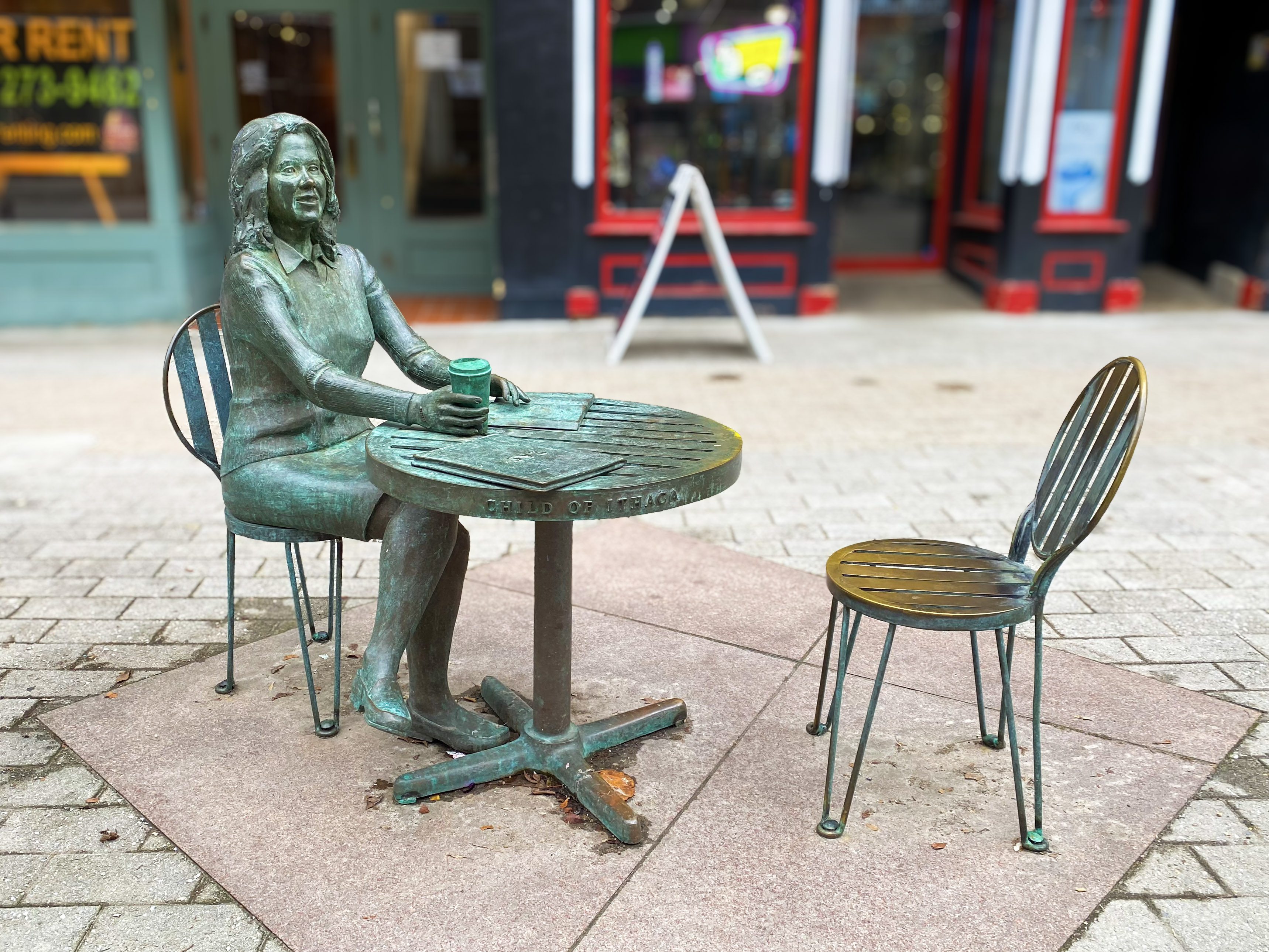}
	\caption{A life-sized sculpture titled \textit{Child of Ithaca} of a young woman sitting at a bistro-style table in Ithaca, New York. The sculpture was created by Roberto G. Bertoia.}
	\Description{A life-sized bronze sculpture of a smiling woman sitting at a bistro-style table, looking up from an open book and holding a cup of coffee. The woman has shoulder-length hair and wears a sweater and skirt. There is one empty chair on the other side of the table. Most of the sculpture has a green patina, aside from brass areas on the empty chair seat and table. On the side of the table, the words “CHILD OF ITHACA” are inscribed in capital letters.}
	\label{fig:sculpture}
\end{figure*}

\textbf{Sculptures} are fully tactile and three-dimensional, and are often made from stone, wood, or metal. They naturally provide tactile feedback to BLV users when trying to access the artwork, and some sculptures may be purposely interactive. The \textit{Child of Ithaca} sculpture, for instance, invites passersby to sit and enjoy some company (Figure \ref{fig:sculpture}). However, similar to mosaics, additional tactile feedback does not necessarily mean that the artwork is more accessible. For example, Anish Kapoor’s \textit{Cloud Gate} (colloquially known as “The Bean”) is a large metal bean-shaped sculpture 33 feet high, 42 feet wide, and 66 feet long. The size makes it such that the additional affordance of being able to touch the surface does not give someone an idea of its shape or scale. 

For public art to be accessible to BLV people, it must convey the content of the art in detail (e.g., the subject, the background, etc.). However, even when descriptions capture the overall visual component of an artwork, there may still be a deeper meaning to the art \textemdash{} such as why the artist chose a specific medium, location-specific connections, and the general history or background inspiring the piece. As such, we encourage artists to provide additional information explaining the artwork’s deeper meaning, further engaging the community and making the art more accessible to all. Accessibility techniques should aim to elucidate these aspects while avoiding overly interpretative or prescriptive outcomes, especially if the deeper understanding is implicit in the original piece \cite{sullivan2020accessibility, cooperhewittguidelines}.

\section{Art Accessibility Techniques}
Prior work on visual art has primarily focused on non-public art, such as art in museums or personal art (e.g., \cite{li2023understanding, rector2017eyes, cavazos2018interactive, cavazos2021accessible, cavazos2021multi, chheda2024engaging}). We first describe work on blind and low vision people’s general experiences with art, then we present existing art accessibility techniques in Table \ref{table:techniques}. We consider which of these techniques are transferable to public spaces and which may not adapt as well. 

\subsection{Existing Techniques}
Blind and low vision people wish to engage with art, just as sighted people do. For example, BLV people hope to engage with visual art in museums for cultural learning, activism purposes, social activities, and more \cite{li2023understanding}. With regards to personal or family art, BLV people felt that having access to a family member’s artwork supported their connection with loved ones \cite{chheda2024engaging}. As such, blind communities, cultural institutions, and researchers have investigated how to make art more accessible. 

In Table \ref{table:techniques}, we present a variety of established practices already implemented in museums as well as more experimental approaches that have been prototyped and evaluated as part of HCI research.

\begin{table*}[ht!]
	\renewcommand{\arraystretch}{1.15}
    \begin{center}
    \caption{Established practices and emerging methods for art accessibility.}
    \Description{Eight access techniques split into established practices and emerging methods. The established practices techniques are: descriptive touch tours, verbal descriptions, and braille and digital versions of print information. The emerging methods are: museum and exhibit accessibility, augmented reality / touchscreen exploration, tactile graphics / 3D models, audio interfaces / multimodal interactions, and AI-supported visual descriptions.}
    \label{table:techniques}
        \begin{tabular}{>{\raggedright}p{0.1\textwidth} | >{\raggedright}p{0.7\textwidth} | >{\raggedright\arraybackslash}p{0.13\textwidth}}
            \toprule
              & \textbf{Access Technique} & \textbf{Examples} \\
            \midrule[\heavyrulewidth]
            Established practices & Descriptive touch tours, during which BLV patrons can physically touch art pieces and docents provide descriptive narration to convey the content and scale of art& \cite{smithsoniantouchtour, tatetouchtour} \\
            \cmidrule{2-3}
             & Verbal descriptions, often provided as narrations through smartphone applications or specific devices& \cite{guggenheimdescriptions, cooperhewittguidelines} \\
            \cmidrule{2-3}
             & Braille and digital versions of print information, such as brochures and maps& \cite{metresources, smithsonianresources} \\
            \midrule
            Emerging methods & Museum and exhibit accessibility, for orienting BLV patrons to where the exhibits and artwork are located& \cite{asakawa2019independent, jain2014pilot, landau2005creating} \\
            \cmidrule{2-3}
             & Augmented reality / touchscreen exploration, allowing users to interact with a digital version of the art (e.g., zoom in, view contours with higher contrast, etc.)& \cite{ahmetovic2021musa, ahmetovic2021touch, guedes2020enhancing, goddard2024seeing} \\
            \cmidrule{2-3}
             & Tactile graphics / 3D models, such as 3D printed, embossed, relief models, or laser cut replicas of artwork& \cite{cavazos2021accessible, shin2020please, holloway2019making, butler2021technology, jansson2003new} \\
            \cmidrule{2-3}
             & Audio interfaces / multimodal interactions, including background music and sonification which can be presented through proxemic audio interfaces &\cite{rector2017eyes, cavazos2021accessible, cavazos2018interactive, luo2023wesee} \\
            \cmidrule{2-3}
             & AI-supported visual descriptions, provided by multimodal large language models such as GPT-4o and applications such as Be My AI or Microsoft Copilot &\cite{aiartdescriptions, bemyai, copilotart, chheda2024engaging} \\
            \bottomrule
        \end{tabular}
    \end{center}
\end{table*}

\subsection{Transferability of Existing Techniques and Design Considerations}
As mentioned above, museums and galleries often feature quiet surroundings, clear pathways for exploring and engaging with the art, and additional information provided by the artist or experts. Furthermore, each work of art in a museum or exhibit is carefully curated to be cohesive with other works, but the often grassroots nature of public art means that multiple pieces of urban artwork may not be quite as harmonious. Since public art often lacks this consistency or predictability, we must consider the transferability of existing art accessibility techniques to urban art scenarios.

\textbf{Learning about urban art.} To make people aware of what art they might encounter in an urban community, there can be two primary approaches: (1) having an accessible repository of existing public art location data, and (2) using visual description tools to identify artwork in real-time. 

For the former, similar to prior efforts on sidewalk accessibility \cite{saha2019project, hara2014tohme}, researchers can leverage Google Street View image data for community members to label public artworks, or create methods for crowdsourcing art locations. Indeed, there exists an online map documenting all known \textit{Lost Cat} graffiti tags in Ithaca, New York \cite{lostcatmap}. The preexisting nature of this resource means that BLV people can refer to it when planning their route. However, we acknowledge that this data may be difficult to compile as there is no single institution responsible for it and the urban landscape is constantly changing. Additionally, given that public art may not always be sanctioned, artists who wish to stay anonymous or who purposefully create art that is hard to find may not wish for their artwork to be indexed in such a way \cite{jiang2024artivism}. 

For more spontaneous identification of public art, AI-powered visual description tools such as Be My AI \cite{aiartdescriptions} can aid in identifying art when someone is already actively traveling in an urban area. Similar to how technologies such as Soundscape \cite{microsoftsoundscape} focus on using spatial audio cues, users can be notified of art pieces in their vicinity through an earcon or a brief one-sentence description of the art, depending on their preference. While automated methods of identifying art in real-time can avoid potential pitfalls with having an out-of-date repository of art locations, this approach may also be more prone to missing art that falls outside of conventional public art expectations (e.g., graffiti art) or art that blends into the surroundings more (e.g., a mural painted using an earth-toned palette).

\textbf{Navigating to urban art.} To make the experience of finding known art locations more accessible, we can build on existing efforts for museum and exhibit accessibility. For example, Asakawa et al. \cite{asakawa2019independent} investigated the efficacy of continuous user location tracking and orientation to provide both navigation assistance and descriptions for artworks within a museum. For urban settings, as one is traveling, they could also receive spatial audio notifications when they are approaching public art. However, navigation and orientation differs between indoor and outdoor settings, and more research is needed to fully understand how this translation could work in practice. 

Furthermore, public artworks will likely be more spread out than works of art in a museum, and will require people to navigate pedestrian infrastructure such as sidewalks and intersections. This presents additional considerations for accessible pathing (e.g., finding resting spots such as benches, avoiding busy intersections). While having information about urban art a priori may make the navigation process easier, it is also important for researchers to consider how to support serendipitous public art discovery for BLV people.

\textbf{Experiencing and understanding urban art.} As for the accessibility of the art itself, verbal descriptions remain as one of the most overarching ways to improve nonvisual access to a work of art. However, it is critical that the descriptions provide enough detail and background, elucidating the public art’s relation to its location in the urban environment, without overly interpreting the artwork for BLV people. While verbal descriptions can be provided by friends and family who might be traveling with BLV people, they may also be crowdsourced \cite{saha2019project, jiang2024artivism, kwon2022supporting}, provided by the artist, or provided by an art expert so that BLV people can appreciate the art independently. 

Tactile engagement may be possible for certain artworks. For example, murals can be converted into tactile graphics, similar to prior work on tactile museum artwork (e.g., \cite{cavazos2021accessible, holloway2019making}). Some mosaics and sculptures may also be accessible for people to touch. However, though sculptures are inherently tactile, they may not be fully accessible. For example, the \textit{Spoonbridge and Cherry} sculpture in Minneapolis, Minnesota is not only too large to be meaningfully interpreted through touch alone, but it is also located within a small pond. For these works that may be indiscernible due to scale or placement in the urban space, technologists could leverage computer vision and 3D printing technology to create handheld models of the art. 

Due to the noisiness of urban surroundings, audio interfaces may not be as effective for accessibility. We encourage researchers to further investigate how combining different techniques can yield more immersive results, building on prior work on multimodal accessibility boosting immersion and engagement \cite{jiang2023beyond}. Having access to narration, light background music, and a 3D model may be more effective at conveying the overarching feeling of a piece compared to narration alone. However, it is also important to avoid sensory overstimulation or overinterpretation to ensure that BLV people can engage with the art in ways that are most comfortable for them.
\section{Conclusion}
Public art is an underexplored area in accessibility, and there are a wealth of interesting future research directions. For example, how do artists navigate tensions between sanctioned vs. unsanctioned art (i.e., graffiti), especially given that part of making public art more accessible involves documenting what this art is and where it is located? On the other hand, how do BLV people wish to engage with these different types of art? Do BLV people wish for public art exploration to emulate their museum visits, or do they seek another experience altogether? 

We also encourage researchers to consider the impact of emerging artificial intelligence technology, especially generative AI models, for providing greater access to the content and context of an artwork. This can take the form of helping artists refine their visual descriptions of a piece of art, generating code files for 3D printing models of artwork, and more. Enhancing the accessibility of public art in urban spaces goes beyond compliance \textemdash{} it enables blind and low vision people to participate in ongoing cultural, political, and social conversations inspired by public artwork. Overall, we advocate for greater emphasis on increasing nonvisual access to public art as a critical component of making urban spaces both enjoyable and accessible.

\section*{Workshop Involvement}
\subsection*{Author Biographies}
Lucy Jiang is a PhD student in Human Centered Design and Engineering at the University of Washington. She is interested in subjective information accessibility, which encompasses art, images, videos, etc., and the accessibility of public spaces. She received her BS in Computer Science from the University of Washington and her MS in Computer Science from Cornell University. In 2024, she was a Cornell PiTech Impact Fellow at the Design Trust for Public Space working on the Neurodiverse City project. 

Jon E. Froehlich is a Professor in the UW Allen School of Computer Science, a Sloan Fellow, NSF CAREER Awardee, and co-founder of Project Sidewalk \textemdash{} a large-scale research initiative to map and assess every sidewalk in the world using Human-AI methods. At UW, Jon directs the Makeability Lab and is Associate Director of CREATE (Center for Research and Education on Accessible Technology and Experiences) and the PacTrans Transportation Center.

Leah Findlater is an Associate Professor in Human Centered Design and Engineering at the University of Washington, where she directs the Inclusive Design Lab and is an Associate Director of CREATE (Center for Research and Education on Accessible Technology and Experiences). Her research is in accessibility and human-centered machine learning.

\subsection*{Rationale for Workshop Attendance}
This workshop paper ties to questions such as, \textit{“how can AI be used to effectively and ethically solve urban accessibility problems and improve the quality of life for all?”} Our work is grounded in our team’s prior experience in urban accessibility, with efforts both in academia (research papers, such as \cite{li2023understanding, saha2019project}) and in practice (VocalEye Crowders, real-world impacts of Project Sidewalk, etc.). Given that a large body of work on urban accessibility for blind and low vision people centers on navigation and orientation, we hope to invite discussion on facets of urban access to improve the enjoyability of a space in addition to its functionality. We are also excited to discuss how artists can get involved in the future as the creators of these artworks and therefore stakeholders in public art accessibility. Lastly, we are interested in learning from other accessibility researchers and practitioners about translating academic research into scalable real-world impact.

\begin{acks}
This work is partially supported by NSF SCC-IRG \#2125087.
\end{acks}

\bibliographystyle{ACM-Reference-Format}
\bibliography{references}

\end{document}